\documentclass[5p,twocolumn]{elsarticle}
\usepackage{ifpdf}
\usepackage[usenames]{color}
\usepackage{graphicx}
\usepackage{amssymb}
\usepackage{amsmath}

\newcommand{\prl}{Phys. Rev. Lett.\ }
\newcommand{\pra}{Phys. Rev. A\ }

\begin{document}

\begin{frontmatter}

\title{Anisotropic sound and shock waves in dipolar Bose-Einstein condensate}

\author[ift,bdu]{P. Muruganandam}
\author[ift]{S. K. Adhikari}

\address[ift]{Instituto de F\'{\i}sica Te\'orica, UNESP - Universidade Estadual Paulista, 01.140-070 S\~ao~Paulo, S\~ao Paulo, Brazil}
\address[bdu]{School of Physics, Bharathidasan University, Palkalaiperur Campus, Tiruchirappalli  620024, Tamilnadu, India}

\begin{abstract}

We study the propagation of anisotropic sound and shock waves in dipolar Bose-Einstein condensate
in three dimensions (3D) as well as in
quasi-two 
(2D, disk shape) and quasi-one (1D, cigar shape) dimensions using the mean-field approach.
 In 3D, the propagation 
of sound and shock waves are distinct in directions parallel and perpendicular to dipole axis with 
the appearance of instability above a critical value corresponding to attraction.
Similar instability appears 
 in 1D and not in 2D. The numerical anisotropic Mach angle agrees with theoretical 
prediction. The numerical sound velocity  in all cases agrees with that calculated from 
Bogoliubov theory. A movie of the anisotropic wave propagation in a dipolar condensate 
is made available as supplementary material. 

\end{abstract}

\end{frontmatter}

\section{Introduction}
 
The alkali metal atoms used in early Bose-Einstein condensate (BEC) experiments have
negligible dipole moment. However, many bosonic atoms
and molecules have large dipole moments and  $^{52}$Cr \cite{pfau} and $^{164}$Dy \cite{dy} BECs, 
with a larger long-range dipolar interaction superposed
on the short-range atomic interaction, have been
realized. Other atoms like  $^{166}$Er \cite{erbium}  
or molecule like $^{7}$Li-$^{133}$Cs \cite{mol}
with even larger dipole moment are candidates
for BEC. The superposition of short-range 
atomic and long-range non-local anisotropic  dipolar interaction  makes the  
study of dipolar BEC (DBEC) very challenging because of the appearance 
of many peculiar properties \cite{pfau,d2007}. 

A tiny object 
can be dragged  in a BEC superfluid below the Landau critical velocity  \cite{landau} without causing any change 
in the superfluid. Above this critical velocity, collective excitations are generated in the BEC
in the form of soliton \cite{science}, turbulent vortex \cite{science,vortex}
 vortex-anti-vortex pair \cite{anisup}, and shock wave \cite{damski}, etc. 
The critical velocity for generation of these diverse
excitations have different values.  It is 0.43 times the sound velocity
for generation of vortices, sound velocity for the sound waves,
1.44 times the sound velocity for solitons \cite{crit}. The original Landau
criterion  \cite{landau} refers to the linear excitations only (phonons and rotons
in He II).

One remarkable property of a DBEC  is its anisotropic superfluidity \cite{anisup}.  
In a dilute DBEC, 
or in a dense
nondipolar BEC, like liquid helium in bulk, sound velocity could be greater than 
 Landau critical velocity due to roton-like  collective excitations \cite{roton}. The Landau critical velocity 
 could be anisotropic  in a DBEC due to anisotropic rotons \cite{anisup}. Also, anisotropic soliton can be 
 generated in a DBEC \cite{dipsol}. Anisotropic collapse has been observed in a $^{52}$Cr  DBEC \cite{anicoll}.
The stability of a 
trapped DBEC \cite{geom}
also shows distinct anisotropy with a disk-shaped trap leading to more 
stability than a cigar-shaped one  \cite{anistab}.

Here we consider   another manifestation of anisotropic 
superfluidity in a DBEC, e.g., anisotropic sound and shock waves.
Shock waves have been widely investigated in different
  systems, such as, on water surface, in supersonic jet and bullet \cite{mach} flights,
in a gas
bubble driven acoustically \cite{sh7}, in a photonic crystal \cite{sh8}, 
in the nonlinear Schr\"odinger 
equation \cite{shnls}
and in BEC of atoms \cite{damski}, and recently observed in BEC 
of polaritons \cite{science}.
Using the mean-field Gross-Pitaevskii  (GP) equation,  we   study the propagation of sound and shock waves 
in a uniform dilute DBEC in three dimensions (3D). 
 Sound propagates anisotropically in a 3D DBEC of small dipole moment 
with 
larger velocity along the axial polarization direction ($z$) and smaller velocity in radial direction  ($\rho \equiv \{x,  y\}$).
Consequently, sound wave emitted from a point has a non-spherical ellipsoid-like front. 
We study the anisotropic (oblique) waves when a tiny object is dragged along the  $z$  and $x$ axes with 
 shock and hypersonic velocities. 
The anisotropic Mach angle  
  \cite{mach} 
for drag along $z$  and $x$ axes
are in agreement with a theoretical prediction. 
For  a critical dipole moment, the sound velocity in $x$-$y$ plane falls to zero and for larger dipole moments
an instability due to attraction begins. 

Next we study the sound propagation in trapped DBEC, which is of interest in  experiments.  
The effect of dipole
moment is more prominent in the cigar (1D) and disk (2D) shapes with strong traps in radial 
and axial  directions. In a spherically-symmetric 3D trap, the effect of dipole moment is less pronounced \cite{anistab}. 
In the 
cigar shape, there is extra attraction due to dipole moment, and   the disk  shape has added repulsion.
We study these 1D \cite{SS} and 2D  \cite{PS} DBEC using reduced GP equations, where the radial and axial 
variables are integrated
out. In the disk-shaped 2D DBEC  in the radial plane, dipole moment 
contribute to extra repulsion and this  makes the system more stable.  Consequently,  sound propagates isotropically 
for all values of dipole moment with a velocity larger than that of a non-dipolar BEC. In the cigar-shaped 1D    
DBEC along the axial direction, dipole moment contribute to extra attraction  and sound propagates with a velocity smaller
than that of a non-dipolar BEC for dipole moment below a critical value. Above this critical value,  
instability appears due to attraction.
The numerical   sound velocities in   1D, 2D and 3D  DBEC are  in agreement 
with Bogoliubov theory. 

\section{Sound and shock waves in Dipolar BEC}

We  study a uniform DBEC using the   
GP
equation  \cite{jb}
 \begin{align}  \label{gp3d} 
i  \frac{\partial \phi_d({\bf r},t)}{\partial t}
   =  \biggr[ -\frac{\nabla_d^2  }{2}   +
 \int \frac{d{\bf  k}}{(2\pi)^d}e^{i{\bf k.r} }  f_d({\bf k}) \biggr] \phi_d({\bf r},t),
\end{align} 
where $f_d({\bf k})=\tilde n_d({\bf k})
 U_d({\bf k}) $, 
$d=1,2,3$ represent 1D,2D, and 3D, respectively, with $  U_d({\bf k })$ the  
 short-range plus dipolar interaction in momentum space and 
 $\tilde n_d({\bf k})\equiv \int
   d{\bf r} e^{-i{\bf k.r}} |\phi_d(\bf r)|^2$  the momentum-space density. The space ({\bf r}) and momentum ({\bf k}) vectors have  $d$ 
   components.
   In 3D, the interaction potential is \cite{jb} 
   \begin{equation} \label{pot3d}
   U_3({\bf k})=4\pi a+ 4\pi a_{dd}\gamma (3\cos^2\theta -1),
   \end{equation}
    with $\cos\theta = k_z/k$, where $\theta$ is the polar angle.
 Here  
$\phi_d({\bf r},t)$  is the 
wave function at time $t$,
 $a$ the atomic scattering length (taken to be positive here),
dipole interaction strength 
 $a_{dd}
=\mu_0 \mu^2 m /(12\pi \hbar^2)$,  
 $\mu$ the (magnetic) dipole moment of an atom, and $\mu_0$ 
the permeability of free space, $1\ge  \gamma \ge-1/2$ is a tuning parameter controlled by 
rotating orienting fields \cite{PS,rotate}. In the following we  use $\gamma=1$ and shall  make some comments on the 
consequence of negative $\gamma$ on sound propagation. 
In Eq. (\ref{gp3d}) length is measured in  units of  $ l_0 \equiv 1$ $\mu$m and
 time $t$
in units of $t_0 \equiv m{ l_0}\-^2/\hbar$, where $m$ is the mass of an atom. 
  
  The quasi-2D shape is achieved with a strong harmonic trap in the axial $z$ direction with oscillator 
  length $d_z=\sqrt{\hbar/m\omega_z}$, where $\omega_z$ is the angular frequency of axial trap.  In this case, 
assuming that the axial excitations are in the oscillator ground state, the $z$ 
  dependence can be integrated out and in Eq. (\ref{gp3d}), we have \cite{PS}
  \begin{equation}\label{pot2d}
  U_2({\bf k})= \frac{4\pi a}{d_z\sqrt{2\pi}}+ \frac{4\pi a_{dd}}{d_z\sqrt{2\pi}} 
  h_{2D}\left(\frac{k_\rho d_z}{\sqrt 2}\right),
\end{equation}
   with
   $h_{2D}(\xi)=2-3 \sqrt \pi \xi e^{\xi^2} {\mbox{erfc}}(\xi)$. 
    The quasi-1D shape is achieved with a strong harmonic trap in  radial  direction with oscillator 
  length $d_\rho=\sqrt{\hbar/m\omega_\rho}$, where $\omega_\rho$ is the angular frequency of radial trap. 
 Then, the $\rho$ 
  dependence can be integrated out to obtain Eq.  (\ref{gp3d}) with \cite{SS}
    \begin{equation}\label{pot1d}
    U_1({\bf k})= \frac{4\pi a}{2\pi d_\rho^2}+ 
\frac{4\pi a_{dd}}{2\pi d_\rho^2}s_{1D}\left(\frac{k_z d_\rho}
    {\sqrt 2}\right),
    \end{equation} 
    with $s_{1D}(\xi)=\int_0^{\infty}e^{-u}
   [3\xi^2/(u+\xi^2)-1]du$.

\subsection{Bogoliubov spectrum}
  
The Bogoliubov spectrum of a uniform gas of density $n_d$ is \cite{anisup,roton}
\begin{equation}\label{bog}
\omega_d ({\bf k})= \sqrt{ ({k^2}/{2})\left[  {k^2}/{2} +
2U_{d}({\bf k}) n_d
\right]}.
\end{equation}
 In 3D, the Bogoliubov velocity  of sound is \cite{anisup,roton}
\begin{equation} \label{3dvel}
c(\theta)=\lim_{|{\bf k}|\to 0} \frac{\omega_3 
({\bf k})}{|{\bf k}|}= c_0\sqrt{1+a_{dd}  (3\cos^2 \theta -1)/a}, 
\end{equation}
where $c_0=\sqrt{4\pi an_3}$ is the sound velocity in nondipolar medium ($a_{dd}=0$) where 
$n_3$ is density in 3D. This velocity is real  for 
$3\cos \theta^2>1$. For $3\cos \theta^2< 1$ $(\theta > 54.7^\circ)$, 
the 3D velocity (\ref{3dvel}) can be 
imaginary for $a_{dd}$ above a critical value,  signaling an instability.

In  2D, the interaction potential 
(\ref{pot2d}) is always positive (repulsive) allowing sound propagation in radial 
direction for all  $a_{dd}$ and the  Bogoliubov sound velocity is
\begin{equation}\label{bsv2d}
c_\rho=\lim_{|{\bf k}|\to 0} \frac{\omega_2 ({\bf k})}{|{\bf k}|}
=\sqrt{2 n_2 \sqrt{2\pi}(a+2 a_{dd})/d_z  } ,
\end{equation}
where $n_2$ is 2D density. If $\gamma=-1/2$ in Eq. (\ref{pot3d}), $c_\rho $ can be imaginary 
corresponding to the attraction instability.

In  1D, the interaction potential  (\ref{pot1d}),
in the  $k_z\to 0$ limit, is negative (attractive) for $a_{dd}>a$.    The Bogoliubov sound velocity is
\begin{equation}\label{bsv1d}
c_z=\lim_{|{\bf k}|\to 0} \frac{\omega_1 ({\bf k})}{|{\bf k}| }
=\sqrt{2 n_1 (a- a_{dd})/d_\rho^2 } ,
\end{equation}
where $n_1$ is 1D density. 
The 1D sound velocity is imaginary for $a_{dd}>a$ when attraction instability appears. 
This instability resulting from  attraction can be avoided by taking $\gamma=-1/2$.

We  also calculated Gaussian variational Lagrangians \cite{gvl} for the 1D and 2D GP equations \cite{jb}
and found  effective interactions $U_d({\bf k})$, which 
 yielded the variational sound velocities identical to   
 the Bogoliubov results (\ref{bsv1d}) and (\ref{bsv2d}).


\subsection{Numerical simulations}

We solve the 1D, 2D, and 3D  GP equations 
using real-time propagation with Crank-Nicolson 
method \cite{CPC} to study sound and shock waves \cite{damski}. 
 In 3D, we consider a generic dipolar atom for numerical simulation with $a= 4$ nm and background 
 density $n_3=100$ $\mu$m$^{-3}\equiv 10^{14}$ cm$^{-3}$.
To study sound propagation, a 3D Gaussian pulse  
is placed at the center of  the  uniform 3D background density  given by 
$\phi_3^2({\bf r},0)=[{100 +40e^{-r^2/(2w^2)}}]$ $\mu$m$^{-3}$,  $w=2$ $\mu$m
subject to 
the  weak expulsive Gaussian potential $V({\bf r})=  0.00001 e^{-r^2/(2w^2)}$ $\mu$m$^{-2}$
at $t=0.$   
On real-time 
evolution of the GP equation, the 3D Gaussian pulse expands. At large 
times a spherical (nondipolar BEC) or an ellipsoid-like (DBEC) sound wave front propagates outwards with 
a central uniform density. From the time evolution of the sound wave front   the anisotropic 
sound velocity in different directions is calculated.

\begin{figure}[!t]
\begin{center}
\includegraphics[width=\linewidth]{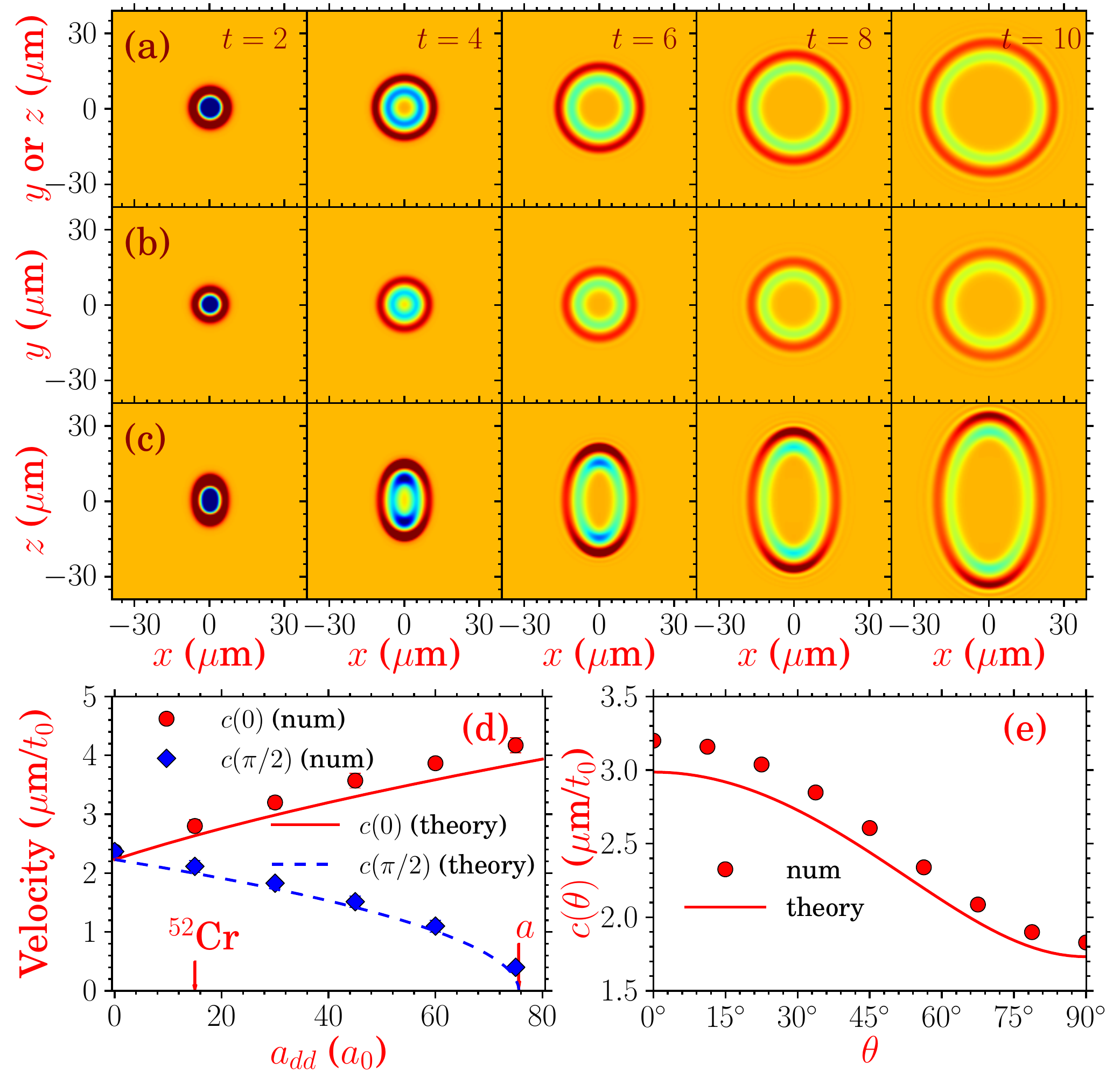}
\end{center}

\caption{ Contour  plot of sound wave pulse (a) $|\phi_3(x,0,z)|^2$ ($a_{dd}=0$), (b)   
$|\phi_3(x,y,0)|^2, (a_{dd}=30a_0)$,
(c) $|\phi_3(x,0,z)|^2, (a_{dd}=30a_0)$, for sound propagation from center outwards at different times 
$t=2,4,6,8,10.$ (d) Numerical (num) and Bogoliubov (theory) sound velocity  for different $a_{dd}$.
(e)  The sound velocity $c(\theta)$ for different polar angle $\theta$. 
}

\label{fig1}
\end{figure}

\begin{figure}[!t]
\begin{center}
\includegraphics[width= \linewidth]{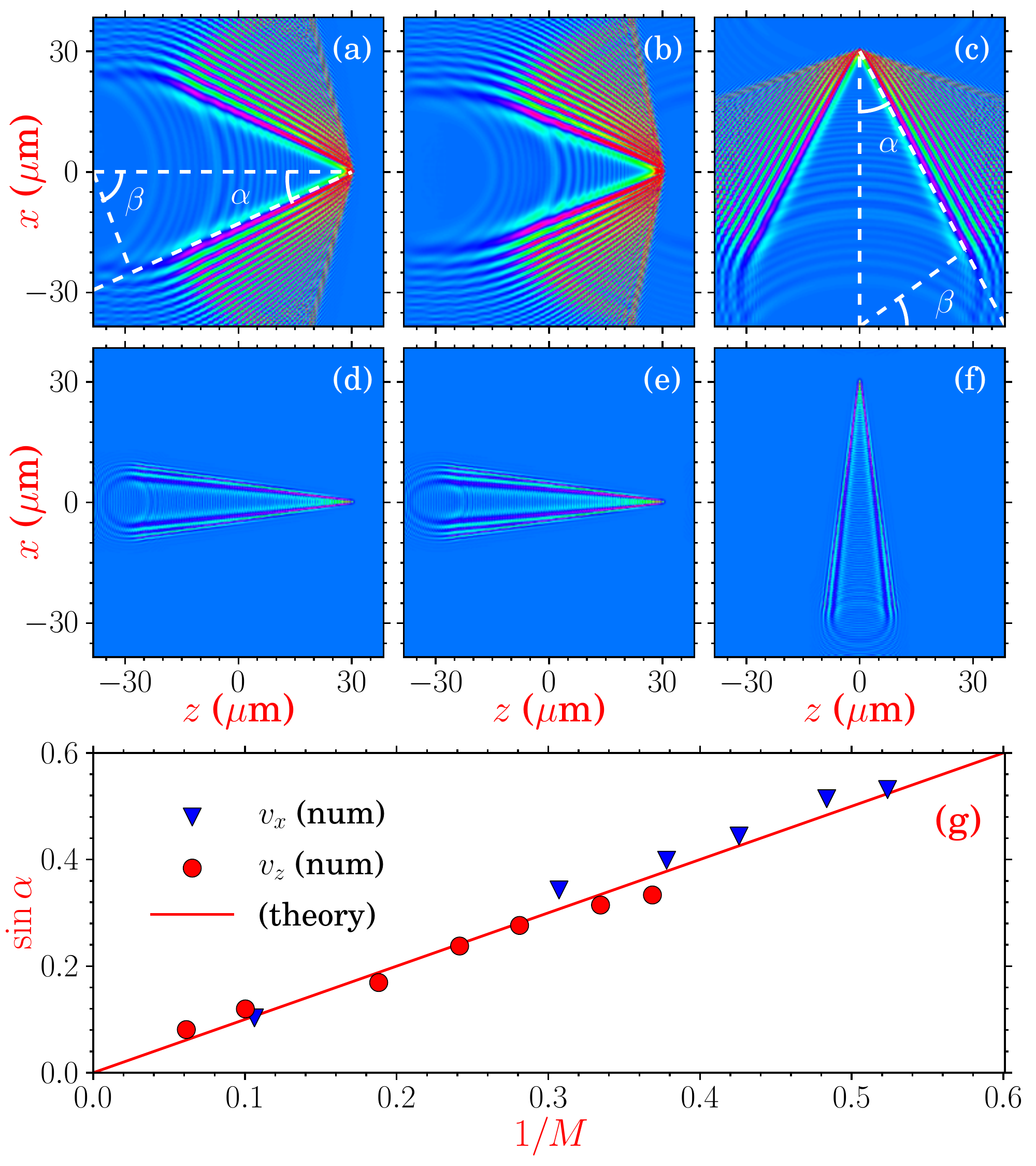}
\end{center}

\caption{  Contour   plot of  $|\phi_3(x,0,z)|^2$ for wave propagation in a 
uniform 3D BEC  for laser  drag  (a) 6 $\mu$m$/{t_0}$ and (d) 30 $\mu$m$/{t_0}$. 
The same in a  DBEC with $a_{dd}=30a_0$ for drag  (b) 6 $\mu$m$/{t_0}$ and (e) 30 $\mu$m$/{t_0}$
along  $z$ axis 
  and for drag  (c) 6 $\mu$m$/{t_0}$ and (f) 30 $\mu$m$/{t_0}$
 along  $x$ axis. 
The angles $\alpha$ and $\beta$   are shown in (a) and (c). 
(g) The $\sin \alpha $ versus $1/M$ plot from theory (\ref{shma})
and numerical simulation (num)
 for different 
drag $v_x$ and $v_z$ 
  along $x$ and $z$ axes in the DBEC. }
\label{fig2}
\end{figure}
 \begin{figure}[!ht]
\begin{center}
\includegraphics[width= \linewidth]{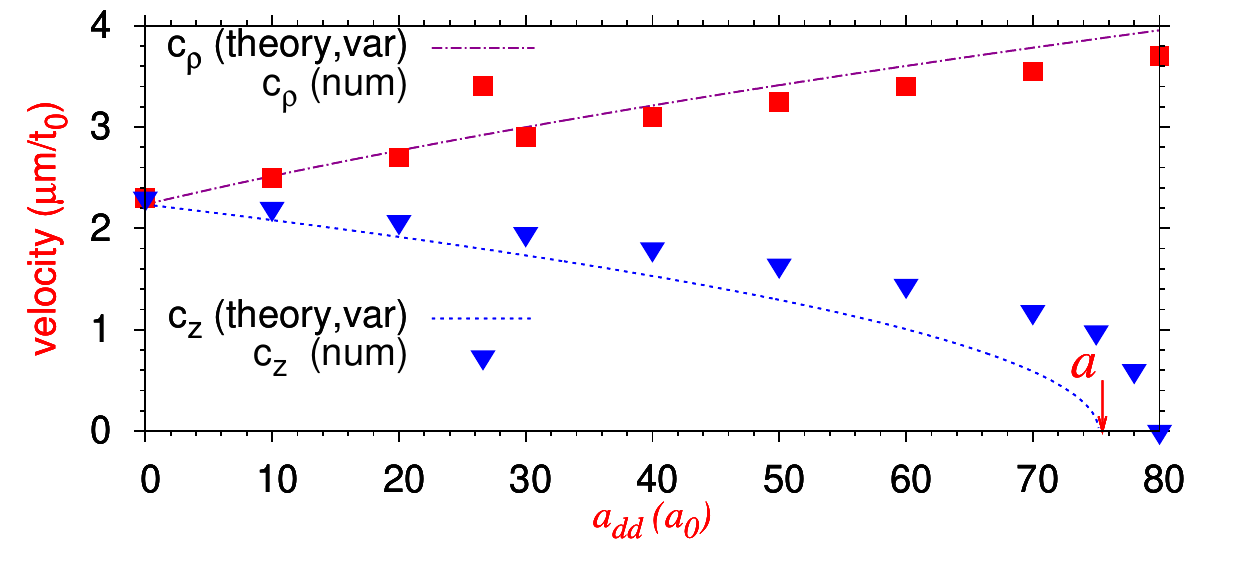}
\end{center}
\caption{ Numerical (num) velocity of sound propagation $c_z$ and $c_\rho$
in 1D and 2D, respectively,  
compared with  Bogoliubov theory
and variational (var) approximation.  
}
\label{fig3}
\end{figure}

Now we study sound propagation in 3D for a uniform DBEC.    In Fig. \ref{fig1} (a), we show isotropic sound propagation in $x$-$z$ plane for a nondipolar BEC. In Fig. \ref{fig1} (b), we show   isotropic sound propagation in $x$-$y$ plane \cite{anisup}
for a DBEC of $a_{dd}=30a_0$.
 In Fig. \ref{fig1} (c), we show anisotropic sound propagation in $x$-$z$ plane for  $a_{dd}=30a_0$. The propagation along $z$ direction 
in a DBEC
has a velocity larger  than that for a nondipolar BEC, whereas 
that in the  $x$-$y$ plane for a DBEC
has a velocity smaller than that for a nondipolar BEC.
The Bogoliubov sound velocity for 
nondipolar atoms is 2.242  $\mu$m$/{t_0}$
compared 
to the numerical velocity of 2.37 $\mu$m$/{t_0}$. For $a_{dd}=30a_0,$  the axial Bogoliubov sound velocity is $c(0)= 3.002
$ $\mu$m$/{t_0}$ (numerical 3.20  $\mu$m$/{t_0}$), and the radial Bogoliubov sound velocity is $c(\pi/2)= 1.741
$ $\mu$m$/{t_0}$ (numerical 1.83  $\mu$m$/{t_0}$).
  In Fig. \ref{fig1} (d), we show the axial and radial sound velocities, $c(0)$ and $c(\pi/2)$, 
versus $a_{dd}$  from numerical simulation and Bogoliubov theory with good agreement between 
the two.  The radial velocity  goes to zero for $a_{dd}=a$ and for
$a_{dd}>a$ attraction instability begins. In Fig. \ref{fig1} (e), we show  sound
velocity $c(\theta)$ versus $\theta$ for $a_{dd}=30a_0$. 

Potentials $
0.2 e^{-[(x-v_xt)^2+z^2]/0.08}$ $\mu$m$^{-2}$ or $0.2 e^{-[x^2+(z-v_zt)^2]/0.08}$ $\mu$m$^{-2}$
of velocity $v_x$ and $v_z$ 
were used to study shock waves \cite{damski} in uniform 3D condensates. These potentials can
be created by moving blue-detuned lasers \cite{anisup}.
These potentials generate 
 waves   in   $x$-$z$ plane  
along $x$ or $z$ direction.  
Typical contour plot in $x$-$z$ plane
of the  3D isotropic oblique wave  is shown in 
Figs. \ref{fig2} (a) and (d) for a   BEC with $v_z= 6$ $\mu$m$/t_0$ (supersonic) and  30  $\mu$m$/t_0$ (hypersonic). 
Anisotropic  waves for a DBEC with $a_{dd}=30a_0$ for $v_x,v_z=6$ $\mu$m$/t_0$ and 30  $\mu$m$/t_0$
are displayed in Figs.  \ref{fig2} (b), (c), (d), and (e).
  The Mach angle $\alpha$ \cite{mach} is related to  laser velocity $v (=v_x,v_z)$ 
and polar angle $ \beta$ of sound propagation  
 by [see, Figs. \ref{fig2} (a) and (c)]
\begin{eqnarray}\label{shma}
\sin \alpha =1/M, \quad M=v/c(\beta).
\end{eqnarray}  
In Figs. \ref{fig2} (a) and (d), $M=6/2.37=2.53$, and $M=30/2.37=12.66$, respectively.
For a DBEC,  sound velocity is different along $x$ and $z$ axes and 
$\alpha $ will be different for laser drag along these axes.
For drag along $x$ axis 
$c(\beta)=c(\alpha) $, and for drag along $z$ axis $c(\beta)=c(\pi/2-\alpha).$
From a full 3D simulation, for drag along $x$ axis, 
in Fig. \ref{fig2} (c),  $\alpha \approx 31^\circ $ and 
$M=v_x/c(\alpha)=6/2.99\approx 2.01$ 
and for drag along $z$ axis, 
in Fig.  \ref{fig2} (b),  $\alpha \approx 18.3^\circ $  and  
$M=v_z/c(\pi/2-\alpha)=6/2.07 \approx  2.90$.
At supersonic velocity, the wave with ripples
outside the Mach angle is oblique, viz. Figs. \ref{fig2} (a) $-$ (c). At hypersonic velocity, the wave, 
mostly confined in the  Mach angle,  is nearly normal,  viz. Figs. \ref{fig2} (d) $-$ (f). 
To test Eq. (\ref{shma})
we did simulation for different $v_x (v_z)$ and 
calculated  $\alpha $ 
 from 2D contour plots   of wave propagation along $x$ and $z$ axes for a
 DBEC with $a_{dd}=30a_0$
and obtained $c(\beta)$ and $M$ using results in Fig. \ref{fig1} (e).   In Fig. \ref{fig2} (g) we plot the theoretical and numerical
results for  $\sin \alpha $ versus $1/M$.
A movie  of the wave of Figs. \ref{fig2} (b) and (c)  is available as
supplementary material \cite{supp} (also available at http://www.youtube.com/watch?v=NR9dB7uZr9M). 
Two independent simulations reported in Figs. \ref{fig1}
and \ref{fig2} confirm the anisotropic nature of sound propagation.

Next we consider sound propagation in trapped medium. The effect of dipolar moment is more pronounced in 
cigar and disk  shapes and we consider these two cases here.  In the cigar shape, dipolar moment leads to 
attraction  which can eventually lead to collapse  for sufficiently large 
dipole moment. In the disk shape the system is always repulsive and propagation of sound is allowed for 
all values of dipole moment. 
The sound wave in cigar and disk shapes are studied by the reduced 1D and 2D GP equations with uniform 
density by putting at the center an initial Gaussian 
pulse on top of an expulsive Gaussian potential at $t=0$ as in 3D. The numerical simulation was done 
with interactions (\ref{pot2d}) and (\ref{pot1d}) using $a=4$ nm and $n_3=10^{14}$ cm$^{-3}$ and $d_z=1$
$\mu$m $
n_2=n_3\sqrt{2\pi}$ and 
$d_\rho=1$ $\mu$m, $n_1=2\pi n_3$ in 2D and 1D, respectively. 
This will lead to the same nondipolar sound velocity in 1D, 2D, and 3D. 
With time evolution, the sound wave propagates outwards from the center. 
In Fig. \ref{fig3}   
sound velocities  calculated for 1D and 2D propagations are plotted  
versus $a_{dd}$ together with the variational and Bogoliubov sound velocities. Instability due to attraction 
 in 1D appears for $a_{dd}
>a$ from Bogoliubov theory and variational approximation. In  numerical simulation 
the instability appears at a slightly 
higher value of $a_{dd}\approx 80a_0\approx 4.23$ nm in 1D.

In this study we took the polarization along axial $z$ axis.  
The conclusion about sound velocity in 1D and 2D will change if we consider the polarization along 
  $x$ axis. The $h_{2D}$ of Eq. (\ref{pot2d}) will be \cite{anisup} $h_{2D}(k_\rho d_z/\sqrt 2)=
-1+$  $3\sqrt \pi k_x^2d_z e^{\xi^2} \mbox{erfc}(\xi)/(\sqrt 2 k_\rho) $ yielding the 2D velocity 
  $c_\rho = \sqrt{2 n_2\sqrt{2\pi}(a-a_{dd})/d_z}$. This velocity will decrease with 
increasing $a_{dd}$ and for $a_{dd}>a$, it becomes imaginary corresponding to attraction.  This 
instability due to attraction
 could be avoided by considering $\gamma=-1/2$ in Eq. (\ref{pot3d}).
Also, for 
polarization along $x$ axis, the 1D velocity $c_z$ will increase with increasing $a_{dd}$  avoiding 
the attraction instability.

\section{Summary and Conclusion}

We   studied anisotropic sound and shock wave propagation in DBEC using the mean-field 
GP equation. In a uniform 3D DBEC, 
sound propagates faster along polarization ($z$) direction than  in radial $x$-$y$  
plane.
For $a_{dd}>a$, there is attraction instability 
in the radial $x$-$y$
plane. 
Oblique waves with distinct Mach angles 
are formed when a tiny object is dragged with  supersonic 
velocity 
 in a DBEC along 
or perpendicular to the polarization direction.
In case of a DBEC, the oblique wave is anisotropic  and the    modified  Mach angle relation 
 (\ref{shma}) is found to hold.  
In a quasi-2D disk-shaped DBEC, sound velocity is found to increase with $a_{dd}$, whereas 
in a quasi-1D cigar-shaped DBEC sound velocity decreases with  $a_{dd}$ and becomes zero for 
 $a_{dd}=a$ and  there is attraction instability for $a_{dd}>a$. 
The  conditions of attraction instability
 can be reverted by tuning $\gamma = -1/2$ in Eq. (\ref{pot3d}).
The  attraction instability   can be implemented by tuning the scattering length 
by
a 
Feshbach resonance \cite{anistab} and/or dipole moment by rotating orienting fields 
\cite{rotate}
and verified 
experimentally.
In all cases the theoretical Bogoliubov sound
velocities are in agreement with numerical simulation.
Experimental confirmation is welcome for  the fascinating anisotropic sound and shock wave propagation 
reported here.

\section*{Acknowledgements}

SKA thanks Luca Salasnich for many discussions on sound waves.
We thank FAPESP and CNPq (Brazil), DST
and CSIR (India) for partial support.

\end{document}